\documentclass[12pt]{article}
\usepackage{epsfig}

\def\beq{\begin{equation}}
\def\eeq#1{\label{#1}\end{equation}}
\def\eeqn{\end{equation}}

\def\beqa{\begin{eqnarray}}
\def\eeqa#1{\label{#1}\end{eqnarray}}
\def\eeqan{\end{eqnarray}}

\let\bar=\overbar

\def\VEV#1{\left\langle{ #1} \right\rangle}

\def\Dslash{\not{\hbox{\kern-4pt $D$}}}
\def\dslash{\not{\hbox{\kern-2pt $\del$}}}

\def\BR{\mbox{\rm BR}}

\def\msb{{\bar{\ssstyle M \kern -1pt S}}}

\def\s#1{\widetilde{#1}}

\newenvironment{Abstract}{\begin{quotation}  }{\end{quotation}}

\def\Title#1{\begin{center} {\Large {\bf #1} } \end{center}}
\def\Acknowledgements{\bigskip  \bigskip \begin{center} \begin{large}
             \bf ACKNOWLEDGEMENTS \end{large}\end{center}}
	     \def\support{\footnote{Speaker on behalf of the \babar\ Collaboration}}

\input{babarsym.tex}

\newcommand{\eqref}[1]{Eq.~(\ref{eq:#1})}

\newcommand{\sweights} {\mbox{$_s{\cal W}eights$}\xspace}

\newcommand{\KstarzI}             {\mbox{$\Kstarz(892)$}}
\newcommand{\KstarI}             {\mbox{$\Kstar(892)$}}
\newcommand{\KstarIp}             {\mbox{$\Kstarp(892)$}}

\newcommand{\rhoz}               {\mbox{$\rho^0$}}

\newcommand{\rhom}               {\mbox{$\rho^-$}}

\newcommand{\fz}                 {\mbox{$f_0$}}
\newcommand{\fI}                 {\mbox{$\fz(980)$}}

\newcommand{\fII}                {\mbox{$f_2(1270)$}}

\newcommand{\fIV}                {\mbox{$f_0(1500)$}}

\newcommand{\fV}                 {\mbox{$f^{\prime}_2(1525)$}}

\begin{document}

\begin{center}
Proceedings of CKM 2012, the 7th International Workshop on the CKM Unitarity Triangle,\\
University of Cincinnati, USA,\\  
28 September - 2 October 2012
\end{center}

\bigskip\bigskip

\Title{Direct \CP\ violation in charmless \B\ decays at \babar}

\bigskip\bigskip


\begin{raggedright}  

{\it Eugenia Maria Teresa Puccio\index{Puccio, E.M.T.}\support\\
Department of Physics\\
Stanford University\\
Stanford, CA, USA.}
\bigskip\bigskip
\end{raggedright}

\begin{Abstract}
	We present \CP\ asymmetry measurements using the full \babar\
	dataset of $471$ million \BB\ pairs for the charmless \B\ decays:
	$\Bz\to\rhoz\Kstarz$, $\Bz\to\fI\Kstarz$, $\Bz\to\rhom\Kstarp$, and
	for the Dalitz plots of $\Bp\to\Kp\Km\Kp$, $\Bp\to\KS\KS\Kp$ and
	$\Bz\to\KS\Kp\Km$.
\end{Abstract}

\section{Introduction}

Charmless \B\ decays are useful probes of the dynamics of weak and strong
interactions. The interference between tree level and penguin contributions
to the same final state can give rise to direct \CP\ violation. The
relative weak phase between tree and penguin amplitudes probes the Unitarity
Triangle angle $\gamma$. In addition to this, enhancement in \CP\
asymmetries with respect to the Standard Model expectations can signify
beyond-the-Standard-Model particle contributions at loop level. 

In this paper, we present the results of analyses conducted by the \babar\
collaboration for the following charmless decay modes:
$\Bz\to\rhoz\Kstarz$, $\Bz\to\fI\Kstarz$, $\Bz\to\rhom\Kstarp$,
$\Bp\to\Kp\Km\Kp$, $\Bp\to\KS\KS\Kp$ and $\Bz\to\KS\Kp\Km$. These analyses
make use of the full \babar\ \FourS\ dataset. This consists of $467$
million \BB\ pairs collected by the \babar\ detector at the \pep2\ \B\
factory, which collides \epem\ asymmetric-energy beams at the \FourS\
resonance~\cite{Aubert:2001tu}. The \B\ meson candidates are characterised
by using two kinematic variables. We take advantage of the precise
kinematic information from the beams to form the variables
$\mes=\sqrt{\frac{s}{4}-p^{*2}_{\B}}$ and
$\DeltaE=E_{\B}^{*}-\frac{\sqrt{s}}{2}$, where
$\left(E^{*}_{\B},\vec{p}^{\,*}_{\B}\right)$ is the \B\ meson four-momentum
in the \epem\ centre-of-mass frame and $\sqrt{s}$ is the centre-of-mass energy.
Signal events are expected to peak around the \B\ mass for \mes\ and around
zero for \DeltaE. There are two main types of backgrounds: the very
abundant \qqbar\ background, where \q\ is either a \u, \d, \s\ or \c\
quark, and the background arising from \B\ decays to other final states. To
distinguish \B\ meson candidates from the continuum \qqbar\ background, 
variables describing the topology of the event are combined in a
multivariate analyser (MVA), such as a neural network or a Fisher
discriminant, in order to maximise their discriminating power.  The
variables \mes, \DeltaE\ and the output of the MVA can either have selection
requirements placed upon them or be supplied as inputs to a maximum
likelihood fit. \BB\ backgrounds are reduced by vetoing events containing a
candidate consistent with a charm particle, and identifying remaining
events by adding a category in the maximum-likelihood fit.
 
\section{Direct \CP\ violation in $\Bz\to\rhoz\Kstarz$, $\Bz\to\fI\Kstarz$
and $\Bz\to\rhom\Kstarp$}

The measurement of direct \CP\ violation in $\Bz\to\rho\Kstar$ can be used
to extract the CKM angle $\gamma$ from $b\to s$ penguin
contributions~\cite{Atwood:2001pf}. However \CP\ asymmetries in these decay
modes have not been previously well measured due to limited statistical
sensitivity and
predictions for these decays are not robust.  \babar\ published
measurements of direct \CP\ violation in $\Bz\to\rhoz\KstarzI$
and $\Bz\to\fz\KstarzI$ based on approximately half the \babar\ dataset. \CP\
asymmetries were found to be consistent with no direct \CP\ violation.
However this analysis revealed only $3.5\sigma$ evidence for the decay
$\Bz\to\fz\KstarzI$, and found no statistically significant rate for
$\Bz\to\rhom\KstarIp$~\cite{Aubert:2006fs}.

Neutral \B\ candidates are reconstructed from their decays to
$\rhoz$ or $\fz$ candidate decaying to $\pip\pim$, or as a $\rhom$ candidate
decaying to $\pim\piz$. These are combined with a $\Kstarz$ or $\Kstarp$
candidates reconstructed as decays to $\Kp\pim$ or $\Kp\piz$,
respectively. The signal region is defined in terms of \mes, \DeltaE\ and a
range in the $K\pi$ invariant mass around the $\KstarI$ mass. To account
for resolution effects due to the \piz\ meson, the $\DeltaE$ signal region
for the $\rhom\Kstarp$ decay is defined to be asymmetric around zero, namely
$-0.17<\DeltaE<0.1\gev$. In addition to a Fisher discriminant formed from
four event-shape variables, continuum background is also suppressed by
applying a selection on the angle between the thrust axis of the \B\
candidate and that of the rest of the event. A main source of \B\ backgrounds
for the $\rhoz/\fz\Kstarz$ decays comes from $\Bz\to\fII\Kstarz$. To
identify the \fII\ contribution in data, a maximum-likelihood fit using \mes,
\DeltaE, the Fisher discriminant and $K\pi$ invariant mass is performed.
\sweights~\cite{Pivk:2004ty} are used to obtain the $\pi\pi$ invariant mass
for the \Kstarz\ signal events.  A fit to the distribution gives a total
of $47\pm3$ \fII\ events in the signal region. 
%

\begin{figure}[!h]
\begin{center}
\includegraphics[height=1.5in]{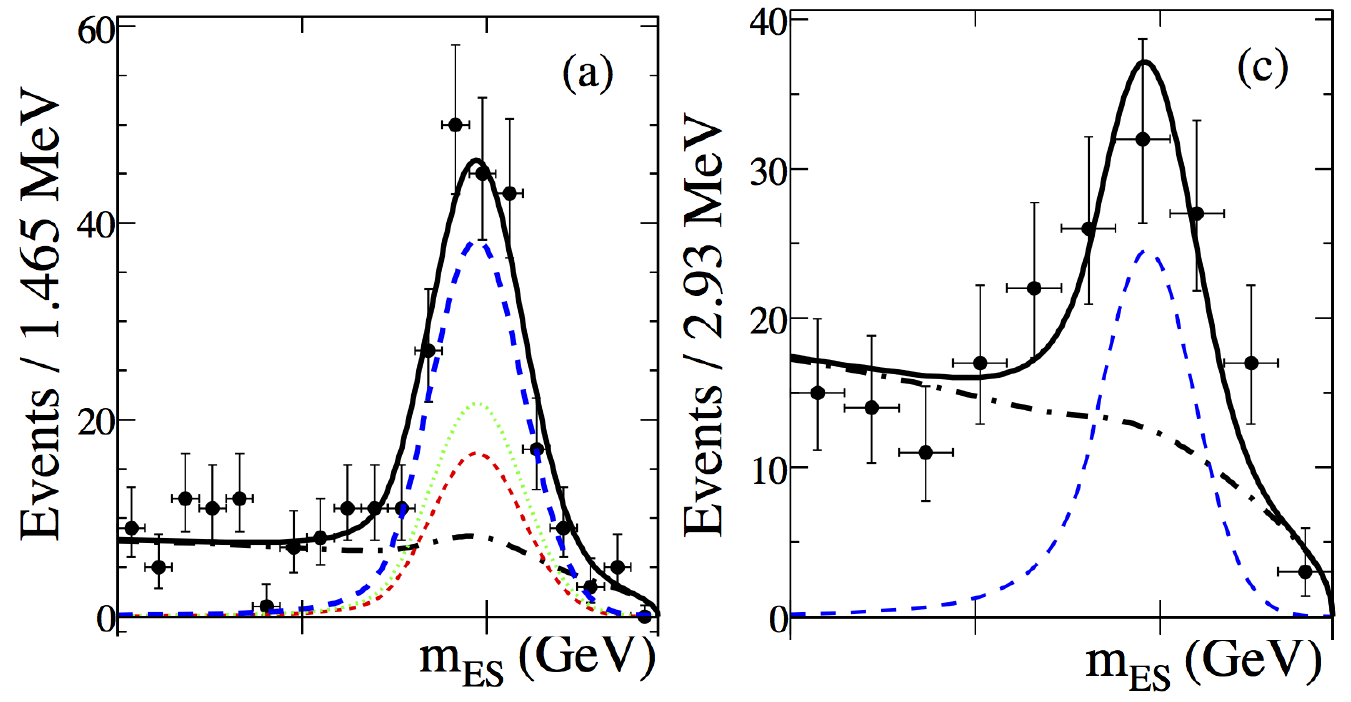}
\includegraphics[height=1.5in]{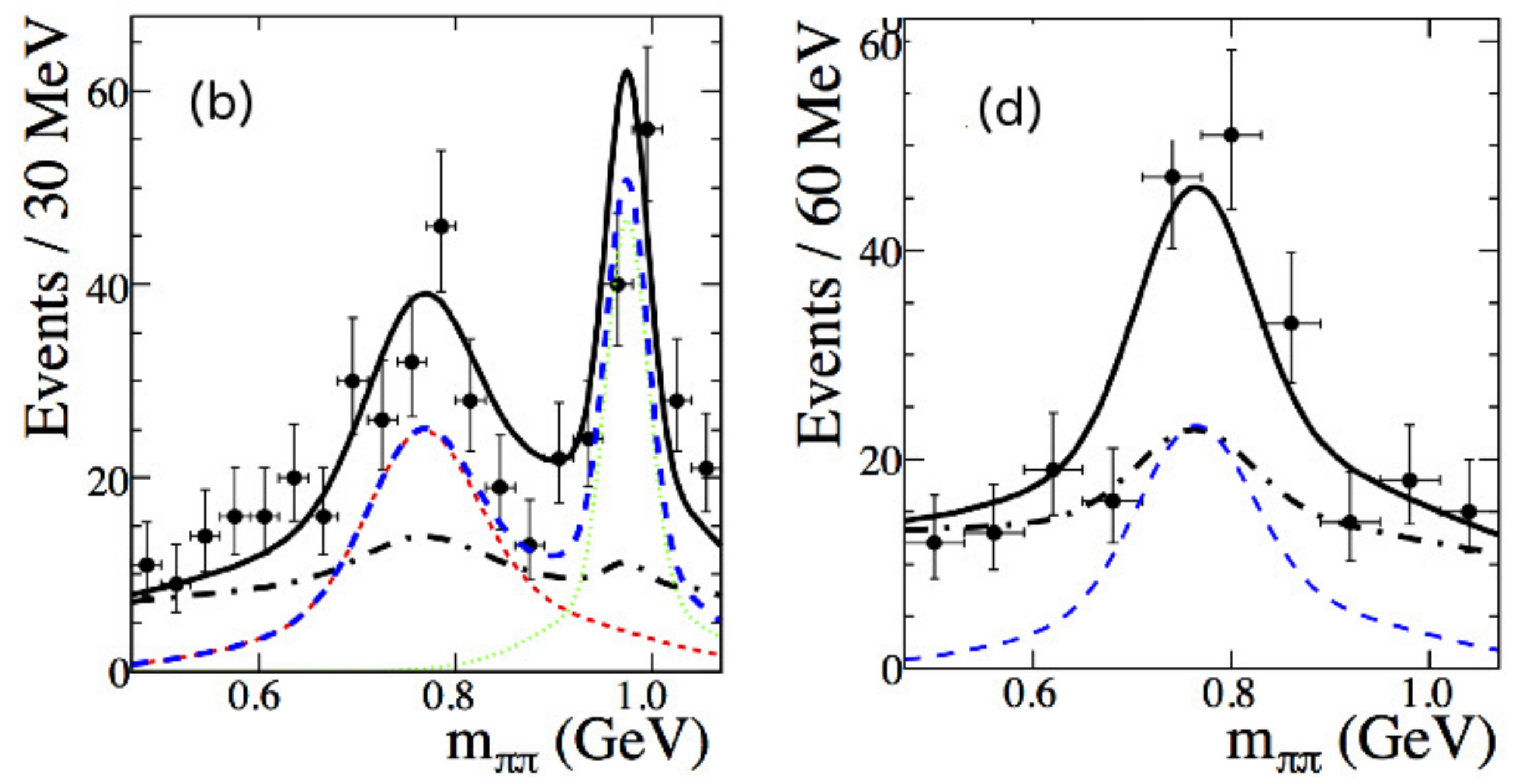}
\caption{\mes\ projections (a,c) and $\pip\pim$ invariant mass projections
	(b,d) for the $\rhoz/\fz\KstarzI$ fit (a,b) and
	the $\rhom\KstarIp$ fit (c,d). The points with error bars show the data
	distribution, the solid curve shows the overall fit result, the
	dot-dashed curve shows total background, and the dashed curve the
	total signal contribution. For the $\Kstarz$ mode, the overall
	signal has been split into a dashed $\rhoz$ component and a dotted
	$\fz$ component. 
}
\label{fig:projection-kstarrho}
\end{center}
\end{figure}

The final maximum-likelihood fit includes seven observables: \mes, \DeltaE,
the Fisher
discriminant, $K\pi$ and $\pip\pim$ invariant masses, and two helicity
angles. \figref{projection-kstarrho} shows the results of the fit to
data. This is the first observation of $\Bz\to\fz\KstarzI$ and
$\Bz\to\rhom\KstarIp$ with significance greater than $5\sigma$. Results for
\CP\ asymmetries are listed in \tabref{Acp-kstar}.  Systematic
uncertainties on \CP\ asymmetries include a potential bias due to the difference in the
interactions of the $\Kp$ and $\Km$ with the detector material
($\approx1\%$). No significant direct \CP\ violation is observed in any of
the three decay modes~\cite{Lees:2011dq}.


\begin{table}[htb]
\begin{center}
\caption{The measured \CP\ asymmetries for the $\rhoz/\fz\Kstarz$ and
$\rhom\Kstarp$ decay modes. }
\label{tab:Acp-kstar}
\begin{tabular}{l|c}  
	Decay mode &  $A_{\CP}$ \\ 
	\hline
	$\Bz\to\rhoz\KstarzI$	& $-0.06\pm 0.09\pm 0.02$	\\
	$\Bz\to\fI\KstarzI$	& $0.07\pm 0.10\pm 0.02$	\\
	$\Bz\to\rhom\KstarIp$	& $0.21\pm 0.15\pm 0.02$	\\
\end{tabular}
\end{center}
\end{table}

\section{Dalitz plot analyses of $\Bp\to\Kp\Km\Kp$, $\Bp\to\KS\KS\Kp$ and
$\Bz\to\KS\Kp\Km$}

\begin{figure}[!htb]
\begin{center}
\includegraphics[height=2.5in]{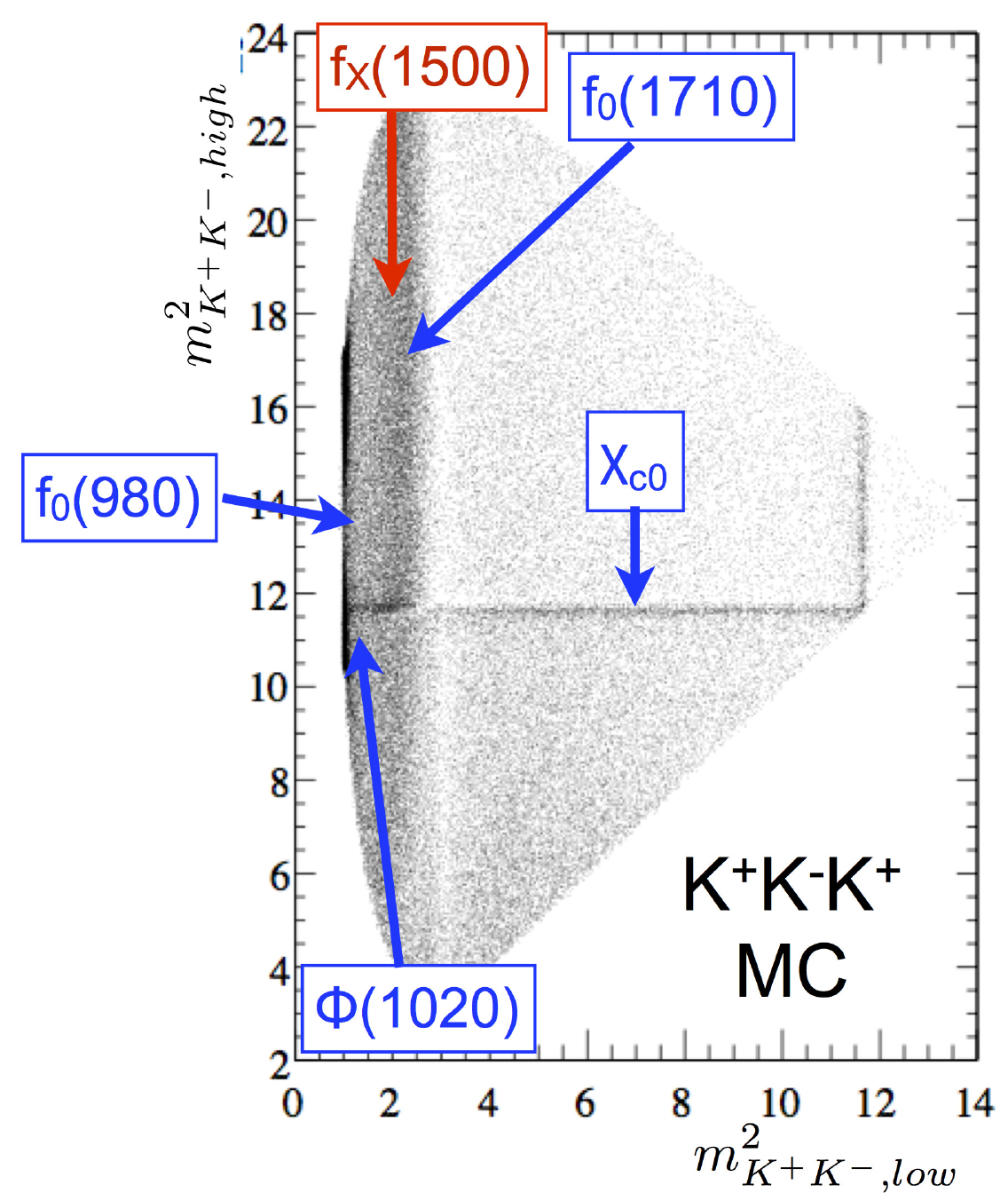}
\caption{ 
	Dalitz plot model from $\Kp\Km\Kp$ MC including an
	$f_{\rm X}(1500)$ component.
}
\label{fig:DP-model-KKK}
\end{center}
\end{figure}

The decays $\Bp\to\Kp\Km\Kp$, $\Bp\to\KS\KS\Kp$ and
$\Bz\to\KS\Kp\Km$ are all $b\to s$ penguin-dominated decay modes. The
inclusive direct \CP\ asymmetries for these decays are predicted to be
small, and in the range $(0.0$~--~$4.7)\%$~\cite{Li:2006jv,Beneke:2003zv}. Any enhancement of the \CP\
asymmetry values would be an indicator of new physics. Previous analyses of
these decay modes have revealed complex Dalitz plot structures that
include a very broad nonresonant contribution, and a poorly understood
scalar resonance, dubbed the $f_{\rm
X}(1500)$~\cite{Aubert:2006nu,Garmash:2004wa,Nakahama:2010nj,Aubert:2007sd}.
An example Dalitz plot Monte-Carlo simulation model for $\Bp\to\Kp\Km\Kp$,
which includes an $f_{\rm X}(1500)$ scalar resonance, is shown in
\figref{DP-model-KKK}.  To create a correct Dalitz plot model, it is
necessary to understand these structures. 

\begin{figure}[!htb]
\begin{center}
\includegraphics[height=2in]{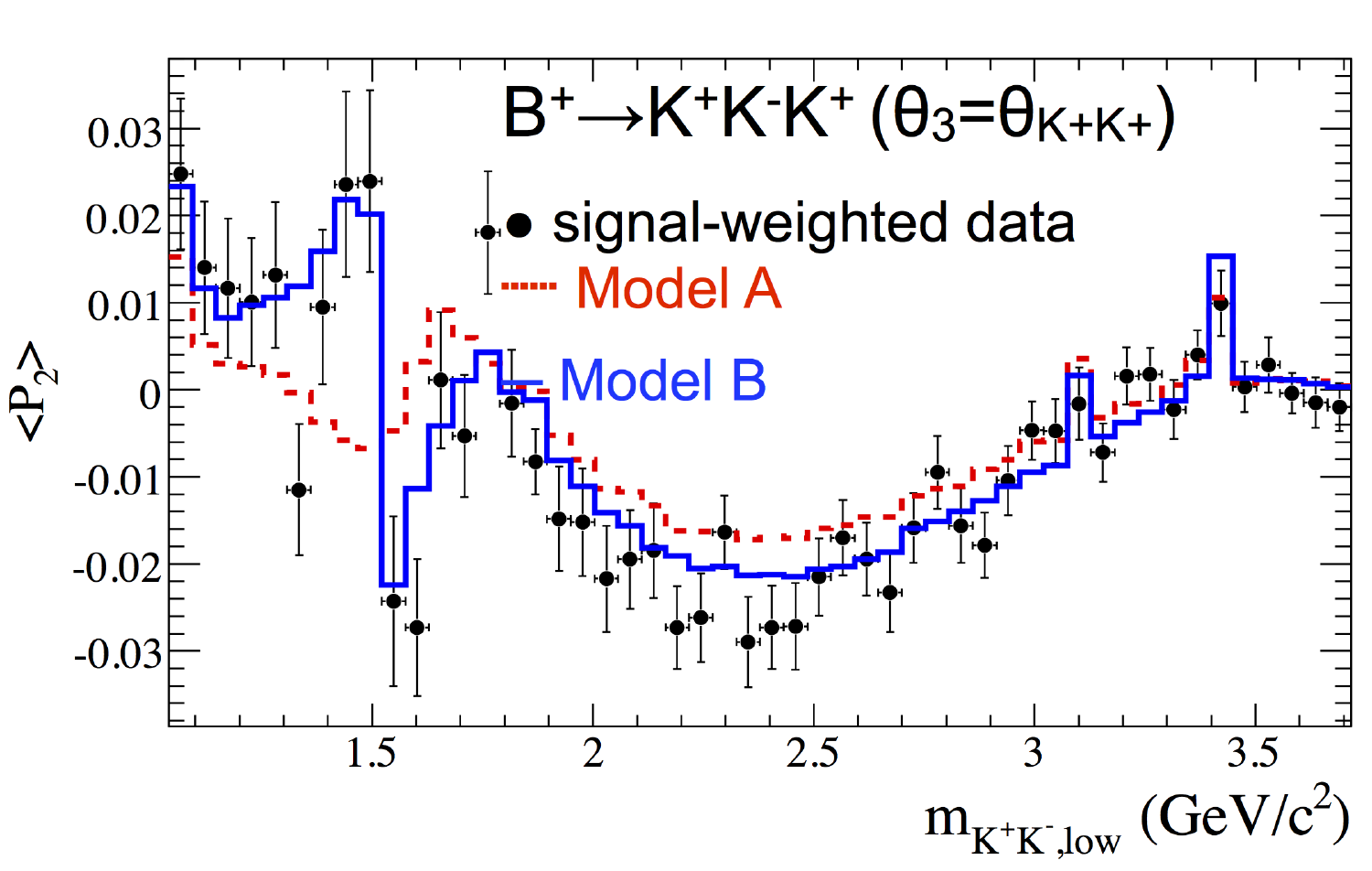}
\caption{
	The $\Bp\to\Kp\Km\Kp$ second-order angular moments for
	signal-weighted data (\sweights) in the region $m_{\Kp\Km\,low}>1.04\gevcc$
	compared to model A (dashed) and model B (solid).
}
\label{fig:moments-KKK}
\end{center}
\end{figure}

Events are reconstructed as three charged-particle tracks consistent with
the kaon hypothesis for $\Bp\to\Kp\Km\Kp$, two charged-particle tracks and a
$\KS\to\pip\pim$ or $\piz\piz$ candidate for $\Bz\to\KS\Kp\Km$, or one
charged-particle track and two $\KS\to\pip\pim$ candidates for $\Bp\to\KS\KS\Kp$.  This
analysis uses Legendre polynomial moments to test the Dalitz plot model of
each $\B\to KKK$ decay mode. Two different models are studied: model A
includes a spin-zero $f_{\rm X}(1500)$ component and model B replaces the
$f_{\rm X}(1500)$ with \fIV\ and \fV\ components. The data are fitted by
varying the mass and width of the $f_{\rm X}(1500)$ and using an
exponential S-wave term for the nonresonant component for model A, and a
polynomial nonresonant model with S-wave and P-wave terms for model B. The
Legendre polynomial moments are defined in terms of $KK$ mass intervals
as:
\begin{equation}
	\VEV{P_{l}(\cos\theta_{3},m_{KK})}\approx\sum_{i=1}^{N}
	P_{l}(\cos\theta^{i}_{3})\,,
\end{equation}
where $\theta_{3}$ is the $KK$ helicity angle, and $N$ is the number of
events in the $m_{KK}$ mass interval. The \sweights\ for signal
moments are then plotted as a function of $KK$ invariant mass, one example
of which is shown in \figref{moments-KKK}, with the distributions for model
A and B overlaid. For all three decay modes, it was found that a model
including the $f_{\rm X}(1500)$ does not describe the data distribution
well. However a significant improvement in log likelihood is achieved when
using model B.
\begin{table}[b]
\begin{center}
\caption{Results for the \CP\ asymmetries for the resonances observed in
the Dalitz plots for $\Bp\to\Kp\Km\Kp$ and $\Bz\to\KS\Kp\Km$.  }
\label{tab:Acp-kkk}
\begin{tabular}{l|c}  
	Decay mode &  $A_{\CP}$ \\ 
	\hline
	$\Bp\to\phi(1020)\Kp$	& $0.13\pm 0.04\pm 0.01$	\\
	$\Bp\to\fI\Kp$		& $-0.08\pm 0.08\pm 0.04$	\\
	$\Bp\to\fV\Kp$		& $0.14\pm 0.10\pm 0.04$	\\
 	$\Bp\to\Kp\Km\Kp$ (NR)	& $0.06\pm 0.04\pm 0.02$	\\ 
	\hline
	$\Bz\to\phi(1020)\KS$	& $-0.05\pm 0.18\pm 0.05$	\\
	$\Bz\to\fI\KS$		& $-0.28\pm 0.24\pm 0.09$	\\
	Other			& $-0.02\pm 0.09\pm 0.03$	\\
\hline
\end{tabular}
\end{center}
\end{table}

\begin{figure}[htb]
\begin{center}
\includegraphics[height=2in]{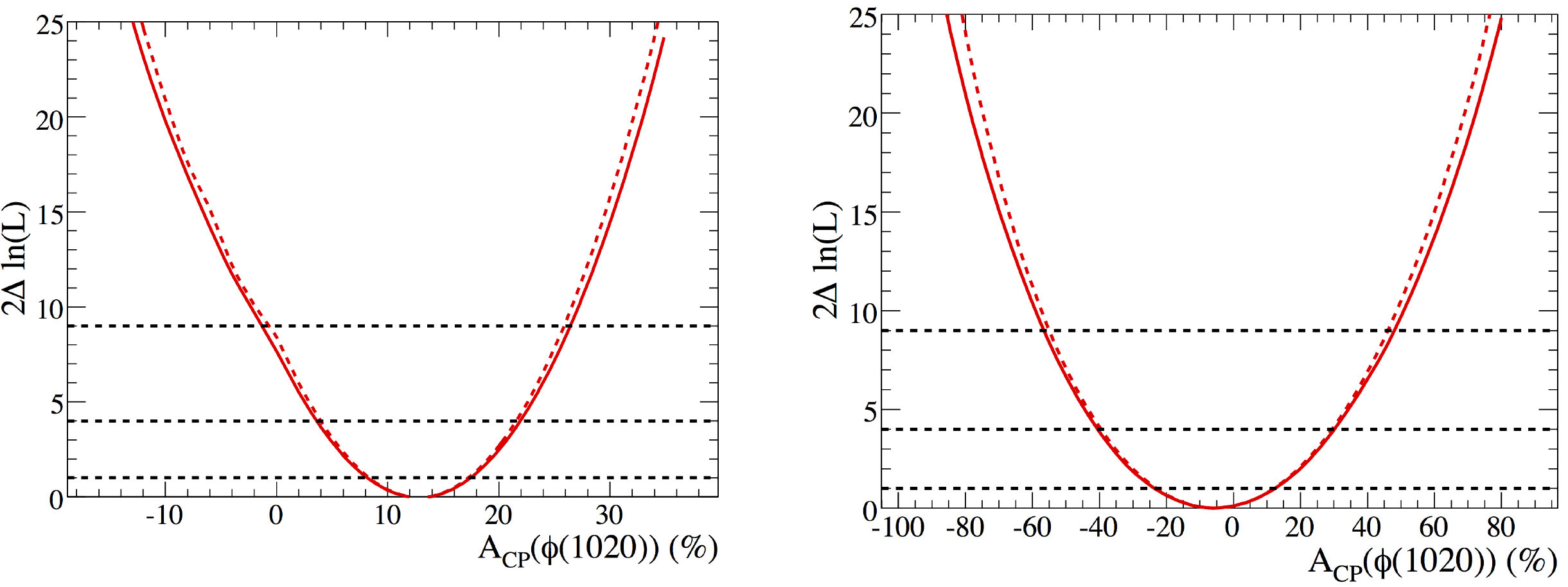}
\caption{
	Scan of the $2\Delta\ln{\cal L}$ with (solid curve) and without
	(dashed curve) systematic uncertainties, as a function of the \CP\
	asymmetry in $\phi(1020)\Kp$ (left) and $\phi(1020)\KS$ (right).
}
\label{fig:log-likelihood-phi}
\end{center}
\end{figure}

\tabref{Acp-kkk} gives the \CP\
asymmetry results for $\Bp\to\Kp\Km\Kp$ and $\Bz\to\KS\Kp\Km$. In the
three-charged-kaon decay mode, the \CP\ asymmetry for $\phi(1020)\Kp$ differs
from zero by about $2.8$ standard deviations. Experimental
precision for the \CP\ asymmetries in $\Bz\to\KS\Kp\Km$ is
however not as good as for $\Bp\to\Kp\Km\Kp$. Log likelihood
plots for the \CP\ asymmetries for $\phi(1020)\Kp$ and $\phi(1020)\KS$ are
shown in \figref{log-likelihood-phi}. The maximum-likelihood fit for
$\Bp\to\KS\KS\Kp$ gives 15 different solutions all within $3\sigma$ of the
global minimum. The amplitudes for resonances in this Dalitz plot are
poorly constrained by the current dataset and therefore only an overall
direct \CP\ asymmetry is quoted, which is consistent with no direct \CP\
violation~\cite{Lees:2012kxa}.

\section{Conclusion}

\babar\ is still actively producing new physics results for charmless
hadronic \B\ decay modes. Most of the direct \CP\ violation results presented
here are consistent with the Standard Model expectations. However
\babar\ measures the \CP\ asymmetry of $\Bp\to\phi(1020)\Kp$ to be higher
than the Standard Model expectation by $2.8\sigma$. This is not yet
evidence of new physics, and more data are required to confirm this
measurement. Both LHCb and Belle 2 will be able to study the Dalitz plot
structures of $\Bp\to\Kp\Km\Kp$ and $\Bz\to\KS\Kp\Km$ with much larger data
samples.

\Acknowledgements{
	I would like to thank Prof. Patricia Burchat, Dr. Bill Dunwoodie,
	Prof. David Brown, Dr. Thomas Latham, Prof. Brian Meadows and the
	\babar\ charmless conveners for the help and suggestions given to
	me in the preparation of this talk. 
}

\def\Discussion{
\setlength{\parskip}{0.3cm}\setlength{\parindent}{0.0cm}
     \bigskip\bigskip      {\Large {\bf Discussion}} \bigskip}
\def\speaker#1{{\bf #1:}\ }
\def\endDiscussion{}



 

\begin{thebibliography}{99}

\bibitem{Aubert:2001tu} 
  B.~Aubert {\it et al.}  (BABAR Collaboration),
  Nucl.\ Instrum.\ Meth.\ A {\bf 479}, 1 (2002).

  \bibitem{Atwood:2001pf} 
  D.~Atwood and A.~Soni,
  Phys.\ Rev.\ D {\bf 65}, 073018 (2002).

  \bibitem{Aubert:2006fs} 
  B.~Aubert {\it et al.}  (BABAR Collaboration),
  Phys.\ Rev.\ Lett.\  {\bf 97}, 201801 (2006).

  \bibitem{Pivk:2004ty} 
  M.~Pivk and F.~R.~Le Diberder,
  Nucl.\ Instrum.\ Meth.\ A {\bf 555}, 356 (2005).
  
  \bibitem{Lees:2011dq} 
  J.~P.~Lees {\it et al.}  (BABAR Collaboration),
  Phys.\ Rev.\ D {\bf 85}, 072005 (2012).

  \bibitem{Li:2006jv} 
  H.~-n.~Li and S.~Mishima,
  Phys.\ Rev.\ D {\bf 74}, 094020 (2006)
  [hep-ph/0608277].

  \bibitem{Beneke:2003zv} 
  M.~Beneke and M.~Neubert,
  Nucl.\ Phys.\ B {\bf 675}, 333 (2003)
  [hep-ph/0308039].

  \bibitem{Aubert:2006nu} 
  B.~Aubert {\it et al.}  (BABAR Collaboration),
  Phys.\ Rev.\ D {\bf 74}, 032003 (2006).

  \bibitem{Garmash:2004wa} 
  A.~Garmash {\it et al.}  (BELLE Collaboration),
  Phys.\ Rev.\ D {\bf 71}, 092003 (2005).

  \bibitem{Nakahama:2010nj} 
  Y.~Nakahama {\it et al.}  (BELLE Collaboration),
  Phys.\ Rev.\ D {\bf 82}, 073011 (2010).

  \bibitem{Aubert:2007sd} 
  B.~Aubert {\it et al.}  (BABAR Collaboration),
  Phys.\ Rev.\ Lett.\  {\bf 99}, 161802 (2007).

  \bibitem{Lees:2012kxa} 
  J.~P.~Lees {\it et al.}  (BABAR Collaboration),
  Phys.\ Rev.\ D {\bf 85}, 112010 (2012).

\end{thebibliography}
\end{document}